\newif\ifTWOCOLUMN\TWOCOLUMNfalse
\TWOCOLUMNtrue

\documentclass[fontsize=9pt,paper=a4,DIV=calc,parskip=never,abstract,captions=tableabove]{scrartcl}
\usepackage[utf8]{inputenc}
\usepackage[T1]{fontenc}
\usepackage{lmodern}
\usepackage{roboto}\renewcommand{\sffamily}{\robotolight}
\usepackage[babel]{microtype}
\usepackage[english]{babel}
\usepackage[autostyle]{csquotes}
\usepackage[iso]{datetime}
\usepackage{calc}
\usepackage{expl3}
\usepackage{currfile}
\usepackage{scrhack}
\usepackage{geometry}

\usepackage[bb=dsserif,cal=dutchcal,frak=euler,scr=pxtx]{mathalpha}
\usepackage{mathtools}
\usepackage{amssymb}
\usepackage{braket}
\usepackage{esint}

\usepackage{array}
\usepackage{booktabs}
\usepackage{multirow}
\usepackage{pbox}
\usepackage[font={sf,footnotesize},labelfont=bf,format=plain,indention=0pt]{caption}
\usepackage{subcaption}
\usepackage{enumitem}
\usepackage[hidelinks,bookmarksnumbered,bookmarks=false]{hyperref}
\usepackage{zref-clever}
\usepackage{zref-xr}
\usepackage[backend=biber,style=phys,biblabel=brackets,sorting=noneyma,pageranges=false,minnames=3,maxnames=6]{biblatex}

\usepackage{graphicx}
\usepackage{xcolor}
\usepackage{tikz,pgfplots,pgfplotstable}

\usetikzlibrary{arrows.meta,calc,external}
\usetikzlibrary{bending,decorations,decorations.markings,fit,intersections,positioning,patterns,patterns.meta,shapes.geometric,shapes.misc}
\pgfplotsset{compat=newest}
\def\dataTitle{}
\def\dataAbstract{}
\def\listAuthors{}
\def\listAffiliations{}
\def\listEmails{}
\newcounter{listAuthors}
\newcounter{listTMP}

\newcommand{\insertTitle}{\dataTitle}
\newcommand{\insertAbstract}{\dataAbstract}
\newcommand{\insertAuthors}{%
	\setcounter{listTMP}{0}%
	\renewcommand*{\do}[1]{%
		\stepcounter{listTMP}%
		\ifnumgreater{\value{listTMP}}{1}{, }{}%
		\ifnumequal{\value{listTMP}}{\value{listAuthors}}{and~}{}%
		##1%
	}%
	\dolistloop{\listAuthors}%
}
\newcommand{\insertAffiliations}{%
	\renewcommand*{\do}[1]{##1\par}%
	\dolistloop{\listAffiliations}%
}
\newcommand{\insertEmails}{%
	\setcounter{listTMP}{0}%
	\renewcommand*{\do}[1]{%
		\stepcounter{listTMP}%
		\ifnumgreater{\value{listTMP}}{1}{, }{}%
		##1%
	}%
	\dolistloop{\listEmails}%
}

\renewcommand{\title}[1]{\def\dataTitle{#1}}
\renewcommand{\abstract}[1]{\def\dataAbstract{#1}}
\newcommand{\authors}[3][]{%
	\ifblank{#1}{}{\listadd{\listEmails}{\href{mailto:#1}{#1}}}%
	\listeadd{\listAuthors}{#2\noexpand\textsuperscript{#3\ifblank{#1}{}{,*}}}%
	\stepcounter{listAuthors}%
}
\newcommand{\affiliations}[2]{%
	\listadd{\listAffiliations}{\textsuperscript{#1}#2}%
}

\makeatletter
\newcommand{\insertHead}{%
	\begin{@twocolumnfalse}\sffamily%
		{\raggedright\huge\bfseries\insertTitle\par}\bigskip%
		{\raggedright\large\bfseries\insertAuthors\par}\medskip%
		{\raggedright\small\setlength{\parskip}{0ex}\insertAffiliations\par\textsuperscript{*}E-mail: \insertEmails\par}%
		\ifx\dataAbstract\empty\else\medskip{\bfseries\insertAbstract\par}\fi%
		\bigskip%
	\end{@twocolumnfalse}%
}

\ifTWOCOLUMN
\else
\fi

\def\@maketitle{%
	\ifTWOCOLUMN%
	\twocolumn[\insertHead]%
	\else%
	\insertHead%
	\fi%
}
\makeatother

\DeclareCiteCommand{\supercite}[\mkbibsuperscript]
{\usebibmacro{cite:init}\bibopenbracket}
{\usebibmacro{citeindex}\usebibmacro{cite:comp}}
{}
{\usebibmacro{cite:dump}\bibclosebracket}

\let\cite=\supercite

\pdfpageattr{/Group << /S /Transparency /I true /CS /DeviceRGB >>}
\let\Re\undefined

\newcommand{\ii}{\mathbb{i}}
\newcommand{\dd}{\mathrm{d}}

\DeclareMathOperator{\Re}{Re}

\DeclareMathOperator*{\argmin}{arg\,min}

\newcommand{\fun}[3][1]{%
	\ifnum #1=0 #2#3                    \else
	\ifnum #1=1 #2\!\left(#3\right)     \else
	\ifnum #1=2 #2\!\left[#3\right]     \else
	\ifnum #1=3 #2\!\left\{#3\right\}   \else
	\ifnum #1=4 #2\!\left<#3\right>     \else
	\fi\fi\fi\fi\fi
}
\renewcommand{\exp}[2][1]{%
	\ifnum #1=0 \,e^{#2}                                \else
	\ifnum #1=1 \operatorname{exp}\!\left(#2\right)     \else
	\ifnum #1=2 \operatorname{exp}\!\left[#2\right]     \else
	\ifnum #1=3 \operatorname{exp}\!\left\{#2\right\}   \else
	\fi\fi\fi\fi
}

\makeatletter
\zref@newprop{compactref}{\compactref{\@currentlabel}{\@currentcounter}}
\zref@addprop{main}{compactref}
\makeatother

\zcsetup{
	ref=compactref,
	typesort={{{othertypes}},figure},
	cap,
	abbrev,
	pairsep  = {,},
	listsep  = {,},
	lastsep  = {,},
	rangesep = {\textendash},
}
\zcRefTypeSetup{equation}{
	refbounds={,(,),},
}

\ExplSyntaxOn
\tl_new:N \g_compactref_tl
\regex_const:Nn \l_compactref_regex {(.+?)([a-zA-Z]*)$}
\NewDocumentCommand{\compactref}{mm}{
	\tl_if_eq:nnTF {#2} {subfigure} {
		\regex_extract_once:NnN \l_compactref_regex {#1} \l_tmpa_seq
		\tl_if_eq:neTF {#1} {\__zrefclever_extract:nnn {\l__zrefclever_type_first_label_tl} {default} {}} {
			#1
		}{
			\tl_if_eq:eeTF {\seq_item:Nn \l_tmpa_seq {2}} {\g_compactref_tl} {
				\seq_item:Nn \l_tmpa_seq {3}
			}{
				\space #1
			}
		}
		\tl_gset:Ne \g_compactref_tl {\seq_item:Nn \l_tmpa_seq {2}}
	}{
		#1
	}
}

\tl_new:N \l_lref_tl
\NewDocumentCommand{\LREF}{mmmm}{
	\tl_set:Nn \l_lref_tl {#4:}
	#1{#2}
	#3{#4}
}
\NewDocumentCommand{\lref}{m}{
	\clist_clear:N \l_tmpa_clist
	\clist_map_inline:nn {#1}{\clist_put_right:Nn \l_tmpa_clist {\l_lref_tl ##1}}
	(\zcref*[noname,ref=subref]{\l_tmpa_clist})
}
\NewDocumentCommand{\Lref}{m}{
	\textbf{\lref{#1}}
}

\NewDocumentCommand{\fref}{mm}{
	\clist_clear:N \l_tmpa_clist
	\clist_map_inline:nn {#2}{\clist_put_right:Nn \l_tmpa_clist {#1:##1}}
	\zcref{\l_tmpa_clist}
}
\NewDocumentCommand{\Fref}{mm}{
	\clist_clear:N \l_tmpa_clist
	\clist_map_inline:nn {#2}{\clist_put_right:Nn \l_tmpa_clist {#1:##1}}
	\zcref[S]{\l_tmpa_clist}
}
\ExplSyntaxOff

\DeclareSortingTemplate{noneyma}{
	\sort{\citeorder}
	\sort{\field{year}}
	\sort{\field{month}}
	\sort{\field{author}}
}

\DeclareBibliographyDriver{entrykeylist}{\mbox{\thefield{entrykey}}|\allowbreak}
\defbibenvironment{entrykeylist}{}{}{}

\BiblatexSplitbibDefernumbersWarningOff

\defbibenvironment{bibliography}{
	\list{\printfield[labelnumberwidth]{labelnumber}}{%
		\setlength{\labelwidth}{\labelnumberwidth}%
		\setlength{\labelsep}{\biblabelsep}%
		\setlength{\leftmargin}{\labelnumberwidth+\biblabelsep+\bibhang}%
		\setlength{\itemindent}{-\bibhang}%
		\setlength{\itemsep}{\bibitemsep}%
		\setlength{\parsep}{\bibparsep}%
	}
	
}{\endlist}{\item}

\newcommand{\weblink}[1]{%
	\iffieldundef{doi}{%
		\iffieldundef{eprint}{%
			\iffieldundef{url}{#1}{\href{\thefield{url}}{#1}}%
		}{\href{https://arxiv.org/abs/\thefield{eprint}}{#1}}%
	}{\href{https://doi.org/\thefield{doi}}{#1}}%
}

\DefineBibliographyStrings{english}{
	and={\&},
	andothers={\textit{et~al\adddot}},
	editor={Ed:},
	in={in}
}

\DeclareNameFormat{author}{%
	\namepartgiveni\addnbspace\namepartfamily%
	\ifthenelse{\value{listcount}<\value{liststop}}%
	{\multinamedelim}%
	{\ifmorenames{\andothersdelim\bibstring{andothers}}{}}%
}
\DeclareNameAlias{editor}{author}

\ExplSyntaxOn
\DeclareFieldFormat{postnote}{
	\ifentrytype{online}{
		\seq_set_split:Nnn \l_tmpa_seq {,} {#1}
		\seq_map_indexed_inline:Nn \l_tmpa_seq {
			\ifnum##1>1,\ \fi\href{\thefield{url}##2}{\texttt{##2}}
		}
	}{#1}
}
\ExplSyntaxOff

\DeclareFieldFormat*{labelnumberwidth}{[#1]}
\DeclareFieldFormat*{title}{\textquote{#1}}
\DeclareFieldFormat*{journaltitle}{\textit{#1}}
\DeclareFieldFormat*{booktitle}{\textit{#1}}
\DeclareFieldFormat*{volume}{\textit{#1}}
\DeclareFieldFormat*{pages}{\mkfirstpage{#1}}
\DeclareFieldFormat*{year}{\textbf{#1}}
\DeclareFieldFormat*{eprint}{(Preprint)\addspace\printfield{eprinttype}:#1}
\DeclareFieldFormat*{type}{\textit{\bibstring{#1}}}

\DeclareBibliographyDriver{article}{%
	\usebibmacro{begentry}\weblink{%
		\printnames{author}\newunit%
		\printfield{title}\newunit%
		\iffieldundef{eprint}{%
			\printfield{journaltitle}\setunit{\addspace}%
			\printfield{year}\newunit%
			\printfield{volume}\newunit%
			\iffieldundef{pages}{\printfield{number}}{\printfield{pages}}%
		}{%
			\printfield{eprint}\newunit
			submitted:\addspace\printfield{month}\addnbspace\printfield{year}%
		}%
	}\usebibmacro{finentry}}

\DeclareBibliographyDriver{book}{%
	\usebibmacro{begentry}\weblink{
		\printnames{author}\newunit%
		\printfield{title}\newunit%
		\printlist{publisher}\newunit%
		\printfield{year}%
	}\usebibmacro{finentry}}

\DeclareBibliographyDriver{incollection}{%
	\usebibmacro{begentry}\weblink{
		\printnames{author}\newunit%
		\printfield{title}\newunit%
		\bibstring{in}\addnbspace\printfield{booktitle}\newunit%
		\printlist{publisher}\newunit%
		\printfield{year}%
	}\usebibmacro{finentry}}

\DeclareBibliographyDriver{thesis}{%
	\usebibmacro{begentry}\weblink{%
		\printnames{author}\newunit%
		\printfield{title}\newunit%
		\printfield{type}\newunit%
		\printlist{institution}\newunit%
		\printfield{year}%
	}\usebibmacro{finentry}}

\DeclareBibliographyDriver{patent}{%
	\usebibmacro{begentry}\weblink{%
		\printnames{author}\newunit%
		\printfield{title}\newunit%
		\iffieldundef{type}{}{\printfield{type}\addnbspace}\printfield{number}\newunit%
		\printfield{year}%
	}\usebibmacro{finentry}}

\DeclareBibliographyDriver{inproceedings}{%
	\usebibmacro{begentry}\weblink{%
		\printnames{author}\newunit%
		\printfield{title}\newunit%
		\bibstring{in}\addnbspace\printfield{booktitle}\newunit%
		\iflistundef{location}{}{\printlist{location}\newunit}%
		\iffieldundef{month}{}{\printfield{month}\addnbspace}\printfield{year}%
		\iffieldundef{note}{}{\addspace(\printfield{note})}%
	}\usebibmacro{finentry}}

\DeclareBibliographyDriver{suppperiodical}{%
	\usebibmacro{begentry}\weblink{%
		Editorial:~\printfield{title}\newunit%
		\iffieldundef{eprint}{%
			\printfield{journaltitle}\setunit{\addspace}%
			\printfield{year}\newunit%
			\printfield{volume}\newunit%
			\iffieldundef{pages}{\printfield{number}}{\printfield{pages}}%
		}{%
			\printfield{eprint}\newunit
			submitted:\addspace\printfield{month}\addnbspace\printfield{year}%
		}%
	}\usebibmacro{finentry}}

\DeclareBibliographyDriver{online}{%
	\usebibmacro{begentry}\weblink{%
		\printnames{editor}\newunit%
		\printfield{title}\newunit%
		\iffieldundef{month}{}{\printfield{month}\addnbspace}\printfield{year}\newunit%
		\printfield{url}\addspace%
		(Accessed:\addnbspace\mkbibdatelong{urlyear}{urlmonth}{})%
	}\usebibmacro{finentry}}

\ifdefined\renewcaptionname
\renewcaptionname{english}{\figurename}{Fig.}
\fi
\DeclareCaptionLabelFormat{vert}{#1~#2~|~}
\tikzset{
	>={Latex[round]},
	font={\footnotesize\sffamily\fontseries{l}\selectfont},
	external/shell escape=-enable-write18,
	external/prefix=figures/,
	/pgf/images/external info,
	png export/.style={
		external/system call/.add={}{&& pdftocairo -singlefile -r 600 -png -transp "\image.pdf" "\image"},
	},
	png export existing/.style={
		external/force remake,
		external/system call={pdftocairo -singlefile -r 600 -png -transp "\image.pdf" "\image"},
	},
	png include/.style={
		/pgf/images/include external/.code={\includegraphics[width=\pgfexternalwidth,height=\pgfexternalheight]{##1.png}},
	},
	label transparent fill/.style n args={2}{ 
		fill=#1,
		fill opacity=0.66,
		text=#2,
		text opacity=1,
		draw=#2,
		draw=none,
	},
	label of style/.style={
		font={\footnotesize\sffamily\bfseries},
		inner sep=1em/3,
		outer sep=0pt,
	},
	label outer of style/.style={
		font={\strut\footnotesize\sffamily\bfseries},
		inner sep=0pt,
		outer sep=0pt,
	},
	label outer north west of/.style={ 
		label outer of style,
		at={(#1.north west)},
		above right=1mm and 0mm,
		align=left,
	},
	label outer north of/.style={
		label outer of style,
		at={(#1.north)},
		above=1mm,
		align=center,
	},
	label outer north east of/.style={
		label outer of style,
		at={(#1.north east)},
		above left=1mm and 0mm,
		align=right,
	},
	label outer west north of/.style={
		label of style,
		at={(#1.north west)},
		below left=0mm and 1mm,
		align=right,
	},
	label north west of/.style={
		label of style,
		at={(#1.north west)},
		below right=2mm and 2mm,
		align=left,
	},
	label north of/.style={
		label of style,
		at={(#1.north)},
		below=2mm,
		align=center,
	},
	label north east of/.style={
		label of style,
		at={(#1.north east)},
		below left=2mm and 2mm,
		align=right,
	},
	label outer east north of/.style={
		label of style,
		at={(#1.north east)},
		below right=0mm and 1mm,
		align=left,
	},
	label outer west of/.style={
		label of style,
		at={(#1.west)},
		left=1mm,
		align=right,
	},
	label west of/.style={
		label of style,
		at={(#1.west)},
		right=2mm,
		align=left,
	},
	label center of/.style={
		label of style,
		at={(#1.center)},
		anchor=center,
		align=center,
	},
	label east of/.style={
		label of style,
		at={(#1.east)},
		left=2mm,
		align=right,
	},
	label outer east of/.style={
		label of style,
		at={(#1.east)},
		right=1mm,
		align=left,
	},
	label outer west south of/.style={
		label of style,
		at={(#1.south west)},
		above left=0mm and 1mm,
		align=right,
	},
	label south west of/.style={
		label of style,
		at={(#1.south west)},
		above right=2mm and 2mm,
		align=left,
	},
	label south of/.style={
		label of style,
		at={(#1.south)},
		above=2mm,
		align=right,
	},
	label south east of/.style={
		label of style,
		at={(#1.south east)},
		above left=2mm and 2mm,
		align=right,
	},
	label outer east south of/.style={
		label of style,
		at={(#1.south east)},
		above right=0mm and 1mm,
		align=left,
	},
	label outer south west of/.style={
		label outer of style,
		at={(#1.south west)},
		below right=1mm and 0mm,
		align=left,
	},
	label outer south of/.style={
		label outer of style,
		at={(#1.south)},
		below=1mm,
		align=left,
	},
	label outer south east of/.style={
		label outer of style,
		at={(#1.south east)},
		below left=1mm and 0mm,
		align=right,
	},
	pics/image/.style n args={4}{
		code={
			\pgfmathsetlength{\pdfpxdimen}{1in/#2}
			\node[inner sep=0pt,outer sep=0pt] (#1) {\includegraphics[#3]{#4}};
		}
	},
	pics/scalebar/.style n args={3}{
		code={
			\matrix[row sep=2pt,inner sep=3pt,outer sep=0pt,cells={inner sep=0pt,outer sep=0pt}]
			{
				\def\tmp{#3}\ifx\tmp\empty\else\node[fill=none,above] (0,0) {#2 #3};\\\fi
				\draw[line width=2pt] ({-(#1*#2)/2},0) -- ({(#1*#2)/2},0);\\
			};
		}
	},
	pics/outline/.style n args={4}{ 
		code={
			\begin{scope}[
				every node/.style={
					outer sep=0pt,
					line width=#1,
				},
				]
				\begin{scope}[transparency group,opacity=0.66]
					\node [color=#2] {\pdfliteral{1 j 1 J 1 Tr}#4};
				\end{scope}
				\node [color=#3] {#4};
			\end{scope}
		}
	},
	tight matrix/.style={
		cells/.expanded={
			inner xsep=\pgfkeysvalueof{/pgf/inner xsep},
			inner ysep=\pgfkeysvalueof{/pgf/inner ysep},
			outer xsep=\pgfkeysvalueof{/pgf/outer xsep},
			outer ysep=\pgfkeysvalueof{/pgf/outer ysep},
		},
		inner sep=0pt,
		outer sep=0pt,
	},
}
\pgfplotsset{
	/pgfplots/plot graphics/includegraphics cmd={\pgfimage},
	unbounded coords=jump,
	every axis plot/.style={semithick},
	every tick/.append style={thin},
	axis line style={semithick},
	tick label style={
		/pgf/number format/fixed,
		/pgf/number format/precision=3,
		/pgf/number format/assume math mode,
	},
	legend cell align=left,
	legend style={
		draw=none,
		fill=none,
		/tikz/every even column/.append style={column sep=1ex},
	},
	legend pos/north/.style={legend style={at={(0.5,1)},below=2mm}},
	legend pos/east/.style ={legend style={at={(1,0.5)},left =2mm}},
	legend pos/south/.style={legend style={at={(0.5,0)},above=2mm}},
	legend pos/west/.style ={legend style={at={(0,0.5)},right=2mm}},
	legend pos/north west/.style={legend style={at={(0,1)},below right=2mm and 2mm}},
	legend pos/north east/.style={legend style={at={(1,1)},below left =2mm and 2mm}},
	legend pos/south west/.style={legend style={at={(0,0)},above right=2mm and 2mm}},
	legend pos/south east/.style={legend style={at={(1,0)},above left =2mm and 2mm}},
	legend pos/outer north/.style={legend style={draw,at={(0.5,1)},above=2mm}},
	legend pos/outer east/.style ={legend style={draw,at={(1,0.5)},right=2mm}},
	legend pos/outer north west/.style={legend style={draw,at={(0,1)},above right=2mm and 0mm}},
	legend pos/outer north east/.style={legend style={draw,at={(1,1)},above left =2mm and 0mm}},
	legend pos/outer east north/.style={legend style={draw,at={(1,1)},below right=0mm and 2mm}},
	legend pos/outer east south/.style={legend style={draw,at={(1,0)},above right=0mm and 2mm}},
	xmin/.forward to=/pgfplots/xtickmin,
	xmax/.forward to=/pgfplots/xtickmax,
	ymin/.forward to=/pgfplots/ytickmin,
	ymax/.forward to=/pgfplots/ytickmax,
	zmin/.forward to=/pgfplots/ztickmin,
	zmax/.forward to=/pgfplots/ztickmax,
	restrict x to limits/.style={
		restrict x to domain=\pgfkeysvalueof{/pgfplots/xmin}:\pgfkeysvalueof{/pgfplots/xmax},
	},
	restrict y to limits/.style={
		restrict y to domain=\pgfkeysvalueof{/pgfplots/ymin}:\pgfkeysvalueof{/pgfplots/ymax},
	},
	restrict meta to limits/.style={
		x filter/.expression={
			(\pgfkeysvalueof{/data point/meta}<\pgfkeysvalueof{/pgfplots/point meta min} ||
			\pgfkeysvalueof{/data point/meta}>\pgfkeysvalueof{/pgfplots/point meta max} ) ? nan : x
		},
	},
	restrict to limits/.style={
		restrict x to limits,
		restrict y to limits,
	},
	imageplot/.style={
		scale only axis,
		axis on top,
		enlargelimits=false,
		tick style={black},
	},
	lineplot/.style={
		scale only axis,
		axis on top,
		enlargelimits={abs value=2mm,auto,true},
		clip marker paths,
	},
	yyaxisA/.style={
		name=#1,
		alias=#1-a,
		lineplot,
		axis y line*=left,
		every axis plot/.append style={Blue},
		cycle list name=linestyles*,
		every y tick/.append style={Blue},
		every outer y axis line/.append style={Blue},
		every axis y label/.append style={Blue},
		every y tick label/.append style={Blue},
		every legend image post/.append style={black},
	},
	yyaxisB/.style={
		name=#1-b,
		anchor=south west,
		at={(#1.south west)},
		lineplot,
		axis x line=none,
		axis y line*=right,
		every axis plot/.append style={Red},
		cycle list name=linestyles*,
		every y tick/.append style={Red},
		every outer y axis line/.append style={Red},
		every axis y label/.append style={Red},
		every y tick label/.append style={Red},
		every legend image post/.append style={black},
	},
	colorbar north/.style 2 args={ 
		colorbar=true,
		colorbar shift/.style={yshift=2mm},
		every colorbar/.style={
			scale only axis,
			axis on top,
			enlargelimits=false,
			at={(parent axis.north west)},
			anchor=south west,
			width/.link={/pgfplots/parent axis width},
			height=2mm,
			xmin/.link={/pgfplots/point meta min},
			xmax/.link={/pgfplots/point meta max},
			ymin=0,
			ymax=1,
			plot graphics/xmin/.link={/pgfplots/point meta min},
			plot graphics/xmax/.link={/pgfplots/point meta max},
			plot graphics/ymin=0,
			plot graphics/ymax=1,
			tick align=outside,
			tick style={black},
			xtick pos=right,
			xticklabel pos=upper,
			ytick=\empty,
			xlabel={#2},
		},
		colorbar/draw/.code={
			\begin{axis}[every colorbar,colorbar shift,colorbar=false]
				\addplot graphics {#1};
			\end{axis}
		},
	},
	colorbar east/.style 2 args={  
		colorbar=true,
		colorbar shift/.style={xshift=2mm},
		every colorbar/.style={
			scale only axis,
			axis on top,
			enlargelimits=false,
			at={(parent axis.south east)},
			anchor=south west,
			width=2mm,
			height/.link={/pgfplots/parent axis height},
			xmin=0,
			xmax=1,
			ymin/.link={/pgfplots/point meta min},
			ymax/.link={/pgfplots/point meta max},
			plot graphics/xmin=0,
			plot graphics/xmax=1,
			plot graphics/ymin/.link={/pgfplots/point meta min},
			plot graphics/ymax/.link={/pgfplots/point meta max},
			tick align=outside,
			tick style={black},
			ytick pos=right,
			yticklabel pos=upper,
			xtick=\empty,
			ylabel={#2},
			plot graphics/lowlevel draw/.code 2 args={%
				\pgfimage[width=####2,height=####1]{\pgfkeysvalueof{/pgfplots/plot graphics/src}}%
			},
			plot graphics/node/.append style={
				anchor=north west,
				rotate=90,
			},
		},
		colorbar/draw/.code={
			\begin{axis}[every colorbar,colorbar shift,colorbar=false]
				\addplot graphics {#1};
			\end{axis}
		},
	},
	colorbar south/.style 2 args={ 
		colorbar north={#1}{#2},
		colorbar shift/.style={yshift=-2mm},
		every colorbar/.append style={
			at={(parent axis.below south west)},
			anchor=north west,
			xtick pos=left,
			xticklabel pos=lower,
		},
	},
	colorbar west/.style 2 args={  
		colorbar east={#1}{#2},
		colorbar shift/.style={xshift=-2mm},
		every colorbar/.append style={
			at={(parent axis.left of south west)},
			anchor=south east,
			ytick pos=left,
			yticklabel pos=lower,
		},
	},
	colorbar log x/.style={
		every colorbar/.append style={
			xmode=log,
			xticklabel={10\textsuperscript{\pgfmathparse{\tick/\logten}\pgfmathprintnumber[assume math mode]{\pgfmathresult}}},
		}
	},
	colorbar log y/.style={
		every colorbar/.append style={
			ymode=log,
			yticklabel={10\textsuperscript{\pgfmathparse{\tick/\logten}\pgfmathprintnumber[assume math mode]{\pgfmathresult}}},
		}
	},
}

\newcommand{\linspace}[4]{
	\pgfmathsetmacro{\valMin}{#1}%
	\pgfmathsetmacro{\valMax}{#2}%
	\pgfmathsetmacro{\valCount}{int(#3)-1}%
	\def#4{}%
	\pgfplotsforeachungrouped \i in {0,...,\valCount}{%
		\pgfmathparse{(\valMax-\valMin)*\i/\valCount+\valMin}%
		\ifx#4\empty%
		\edef#4{\pgfmathresult}%
		\else%
		\edef#4{#4,\pgfmathresult}%
		\fi%
	}%
}

\newcounter{InterpRowCount}
\newcommand{\interp}[4]{
	\edef\queryColumn{#2}%
	\edef\queryValues{#3}%
	\setcounter{InterpRowCount}{0}%
	\foreach \x in #3{\stepcounter{InterpRowCount}}%
	\pgfmathtruncatemacro{\outputRows}{\theInterpRowCount}%
	\pgfplotstablenew[
	columns=\queryColumn,
	create on use/\queryColumn/.style={create col/set list/.expanded={\queryValues}}
	]{\outputRows}{#4}%
	\pgfplotstableforeachcolumn{#1}\as\col{%
		\ifx\col\queryColumn\else%
		\pgfplotstablecreatecol[create col/set={nan}]{\col}{#4}%
		\fi%
	}%
	\pgfplotstablesort[sort key={\queryColumn}]{#1}{#1}%
	\pgfplotstableforeachcolumnelement{\queryColumn}\of#1\as\inputXMax{%
		\edef\inputRow{\pgfplotstablerow}%
		\ifnum\inputRow>0%
		\pgfmathfloatparsenumber{\inputXMin}%
		\pgfmathfloatgetflagstomacro{\pgfmathresult}{\inputXMinFlags}%
		\pgfmathfloatparsenumber{\inputXMax}%
		\pgfmathfloatgetflagstomacro{\pgfmathresult}{\inputXMaxFlags}%
		\pgfmathtruncatemacro{\isTrue}{\inputXMinFlags<3 && \inputXMaxFlags<3}%
		\ifnum\isTrue=1%
		\pgfplotstableforeachcolumnelement{\queryColumn}\of#4\as\outputX{%
			\pgfmathtruncatemacro{\isTrue}{\inputXMin<=\outputX && \outputX<=\inputXMax}%
			\ifnum\isTrue=1%
			\edef\outputRow{\pgfplotstablerow}%
			\pgfplotstableforeachcolumn{#4}\as\col{%
				\ifx\col\queryColumn\else%
				\pgfplotstablemodifyeachcolumnelement{\col}\of#4\as\cell{%
					\ifnum\pgfplotstablerow=\outputRow%
					\pgfplotstablegetelem{\numexpr\inputRow-1\relax}{\col}\of#1%
					\edef\inputYMin{\pgfplotsretval}%
					\pgfplotstablegetelem{\inputRow}{\col}\of#1%
					\edef\inputYMax{\pgfplotsretval}%
					\pgfmathfloatparsenumber{\inputYMin}%
					\pgfmathfloatgetflagstomacro{\pgfmathresult}{\inputYMinFlags}%
					\pgfmathfloatparsenumber{\inputYMax}%
					\pgfmathfloatgetflagstomacro{\pgfmathresult}{\inputYMaxFlags}%
					\pgfmathtruncatemacro{\isTrue}{\inputYMinFlags<3 && \inputYMaxFlags<3}%
					\ifnum\isTrue=1%
					\pgfmathsetmacro{\cell}{(\inputYMax-\inputYMin)/(\inputXMax-\inputXMin)*(\outputX-\inputXMin)+\inputYMin}%
					\fi%
					\fi%
				}%
				\fi%
			}%
			\fi%
		}%
		\fi%
		\fi%
		\edef\inputXMin{\inputXMax}%
	}%
}

\newcommand{\wrapOverNan}[4]{
	\edef\xColumn{#2}%
	\edef\yColumn{#3}%
	\def\outputList{}%
	\def\outputRows{0}%
	\pgfplotstableforeachcolumnelement{\yColumn}\of#1\as\yThis{%
		\edef\inputRow{\pgfplotstablerow}%
		\ifnum\inputRow>0%
		\pgfmathtruncatemacro{\isSkip}{abs(\yThis-\yPrev)>pi*(1+0.1)}
		\ifnum\isSkip=1%
		\pgfplotstablegetelem{\numexpr\inputRow-1\relax}{\xColumn}\of#1%
		\edef\xPrev{\pgfplotsretval}%
		\pgfplotstablegetelem{\inputRow}{\xColumn}\of#1%
		\edef\xThis{\pgfplotsretval}%
		\pgfmathparse{(\xThis+\xPrev)/2}%
		\ifx\outputList\empty%
		\edef\outputList{\pgfmathresult}%
		\else%
		\edef\outputList{\outputList,\pgfmathresult}%
		\fi%
		\pgfmathtruncatemacro{\outputRows}{\outputRows+1}%
		\fi%
		\fi%
		\edef\yPrev{\yThis}%
	}%
	\pgfplotstablenew[
	columns={\xColumn,\yColumn},
	create on use/\xColumn/.style={create col/set list/.expanded={\outputList}},
	create on use/\yColumn/.style={create col/set={nan}}
	]{\outputRows}{#4}%
	\pgfplotstablevertcat{#4}{#1}%
	\pgfplotstablesort[sort key={\xColumn}]{#4}{#4}%
}

\definecolor{Red}      {RGB}{204,  83,  77}
\definecolor{Orange}   {RGB}{222, 143,  68}
\definecolor{Yellow}   {RGB}{230, 203,  81}
\definecolor{Lime}     {RGB}{135, 174,  52}
\definecolor{Green}    {RGB}{  7, 141,  58}
\definecolor{Teal}     {RGB}{  0, 100,  83}
\definecolor{Cyan}     {RGB}{  0, 150, 164}
\definecolor{Azure}    {RGB}{ 84, 200, 253}
\definecolor{Blue}     {RGB}{  0, 146, 246}
\definecolor{Purple}   {RGB}{ 91,  93, 233}
\definecolor{Magenta}  {RGB}{179, 118, 181}
\definecolor{Pink}     {RGB}{130,  77,  97}

\definecolor{Red_1}    {RGB}{205, 112, 103}
\definecolor{Orange_1} {RGB}{224, 163, 111}
\definecolor{Yellow_1} {RGB}{235, 213, 135}
\definecolor{Lime_1}   {RGB}{160, 185, 102}
\definecolor{Green_1}  {RGB}{ 77, 154,  90}
\definecolor{Teal_1}   {RGB}{ 54, 116, 102}
\definecolor{Cyan_1}   {RGB}{ 89, 165, 175}
\definecolor{Azure_1}  {RGB}{148, 210, 245}
\definecolor{Blue_1}   {RGB}{106, 161, 234}
\definecolor{Purple_1} {RGB}{122, 113, 220}
\definecolor{Magenta_1}{RGB}{187, 141, 188}
\definecolor{Pink_1}   {RGB}{142,  98, 114}

\definecolor{Red_2}    {RGB}{208, 137, 129}
\definecolor{Orange_2} {RGB}{227, 182, 146}
\definecolor{Yellow_2} {RGB}{239, 223, 175}
\definecolor{Lime_2}   {RGB}{181, 197, 141}
\definecolor{Green_2}  {RGB}{113, 168, 119}
\definecolor{Teal_2}   {RGB}{ 86, 133, 121}
\definecolor{Cyan_2}   {RGB}{130, 180, 187}
\definecolor{Azure_2}  {RGB}{186, 221, 243}
\definecolor{Blue_2}   {RGB}{144, 176, 229}
\definecolor{Purple_2} {RGB}{144, 133, 214}
\definecolor{Magenta_2}{RGB}{197, 164, 197}
\definecolor{Pink_2}   {RGB}{155, 120, 132}

\definecolor{Red_3}    {RGB}{212, 160, 153}
\definecolor{Orange_3} {RGB}{231, 201, 177}
\definecolor{Yellow_3} {RGB}{243, 233, 206}
\definecolor{Lime_3}   {RGB}{201, 210, 174}
\definecolor{Green_3}  {RGB}{144, 182, 147}
\definecolor{Teal_3}   {RGB}{115, 150, 140}
\definecolor{Cyan_3}   {RGB}{164, 197, 202}
\definecolor{Azure_3}  {RGB}{215, 233, 245}
\definecolor{Blue_3}   {RGB}{174, 193, 229}
\definecolor{Purple_3} {RGB}{163, 152, 212}
\definecolor{Magenta_3}{RGB}{208, 186, 208}
\definecolor{Pink_3}   {RGB}{170, 142, 151}

\pgfplotsset{
	short line legend/.style={
		legend image code/.code={
			\draw[##1] plot coordinates {(-1.5mm,0mm) (1.5mm,0mm)};
		},
	},
	short mark legend/.style={
		legend image code/.code={
			\draw[only marks,##1] plot coordinates {(0mm,0mm)};
		},
	},
	short area legend/.style={
		legend image code/.code={
			\fill[draw=none,##1] (-1.5mm,-1.5mm) rectangle (1.5mm,1.5mm);
		},
	},
	short double area legend/.style n args={2}{
		legend image code/.code={
			\fill[draw=none,##1,#1] (-1.5mm,-1.5mm) -- (1.5mm,1.5mm) -- (-1.5mm, 1.5mm);
			\fill[draw=none,##1,#2] (-1.5mm,-1.5mm) -- (1.5mm,1.5mm) -- ( 1.5mm,-1.5mm);
		},
	},
	short frame legend/.style={
		legend image code/.code={
			\fill[##1] (-1.5mm,-1.5mm) rectangle (1.5mm,1.5mm);
		},
	},
}

\def\SiO{SiO\textsubscript{2}}
\def\TiO{TiO\textsubscript{2}}
\def\BTO{BaTiO\textsubscript{3}}
\def\LNO{LiNbO\textsubscript{3}}

\title{Hybrid Barium Titanate Waveguide Designs For Efficient Nonlinear Frequency Conversion}

\authors{Trevor G.\ Vrckovnik}{1,2,3}
\authors{Dennis Arslan}{1}
\authors{Falk Eilenberger}{1,2,3}
\authors[sebastian.wolfgang.schmitt@iof.fraunhofer.de]{Sebastian W.\ Schmitt}{1,2}

\affiliations{1}{Fraunhofer Institute for Applied Optics and Precision Engineering IOF, Albert-Einstein-Str.\ 7, 07745 Jena, Germany}
\affiliations{2}{Institute of Applied Physics, Abbe Center of Photonics, Friedrich Schiller University Jena, Albert-Einstein-Str.\ 15, 07745 Jena, Germany}
\affiliations{3}{Max Planck School of Photonics, Albert-Einstein-Str.\ 15, 07745 Jena, Germany}

\abstract{Barium titanate (\BTO) is emerging as a powerful integrated photonic material, combining strong $\boldsymbol{\chi}^{(2)}$ and electro-optic nonlinearities with rapidly improving thin-film waveguide quality. Recent demonstrations of low-loss \BTO\ waveguides and high-Q resonators have established \BTO-on-insulator as a promising platform for next-generation frequency-conversion and quantum photonic technologies. However, while \BTO\ electro-optic modulators are now well developed, nonlinear \BTO\ waveguide engineering remains comparatively immature. Techniques widely used in lithium niobate, such as periodic poling for quasi-phase-matching, are poorly suited to \BTO\ because epitaxial thin films exhibit high coercive fields, strong strain-clamping effects, multivariant domain structures, and slow, complex switching dynamics. These factors make accurate periodic poling challenging and hinder the development of efficient $\boldsymbol{\chi}^{(2)}$ frequency converters. Here, we introduce a fabrication-robust alternative based on linear-nonlinear hybrid waveguides, where \TiO\ is selectively incorporated into \BTO\ ridge waveguides to enhance nonlinear mode overlap while relying solely on modal phase-matching. Using coupled-mode-theory simulations, we identify phase-matched geometries and show that the hybrid design achieves a 2.75x increase in normalized second harmonic generation efficiency over monolithic \BTO\ waveguides. The uniform, lithographically defined cross-section makes the approach highly scalable. These results position hybrid \BTO-\TiO\ waveguides as a practical route to CMOS-compatible, high-efficiency $\boldsymbol{\chi}^{(2)}$ devices for integrated quantum photonics.} 

\begin{document}
\maketitle

\section{Introduction}
Thin-film ferroelectric oxides have recently attracted substantial interest for integrated photonics because they combine strong nonlinear and electro-optic properties with compatibility to semiconductor fabrication workflows \cite{sandoEpitaxialFerroelectricOxide2018,demkovSiintegratedFerroelectricsPhotonics2022,alexanderManufacturablePlatformPhotonic2025}. Among these materials, barium titanate (\BTO) stands out due to its large Pockels coefficient, high second-order nonlinear susceptibility ($\boldsymbol{\chi}^{(2)}$), and ability to form epitaxial layers on silicon-based substrates \cite{Karvounis202011}. Recently, advances in thin-film growth, planarization, and patterning have enabled low-loss \BTO\ waveguides and high-Q resonators from \BTO-on-insulator platforms \cite{Riedhauser202501,Raju202504,Kim20250723}.

The strong electro-optic effect of \BTO\ has enabled compact modulators with low drive voltages and multi-GHz bandwidths \cite{Abel201811,Dong202311,Xiong201402,Chelladurai202503}, and non-volatile phase shifters \cite{Abel201811}. The same large $\boldsymbol{\chi}^{(2)}$ nonlinearity suggests strong potential for frequency conversion and photon-pair generation processes such as second harmonic generation (SHG). However, in contrast to lithium niobate (\LNO) where quasi-phase-matching via periodic poling underpins a large ecosystem of nonlinear devices \cite{Wang201811,Jankowski202001,Zhao202004}, it remains questionable whether \BTO\ thin films support reliable and scalable domain inversion engineering. Strong substrate-induced clamping stabilizes multivariant, strain-dependent domain configurations and coercive fields, while the reduced remnant polarization of \BTO\ thin films promotes relaxation and back-switching, together limiting controlled and repeatable 180° domain switching\cite{palSubsecondOpticallyControlled2025,everhardtFerroelectricDomainStructures2016,liEvolutionEpitaxialBaTiO32022,choudhuryStrainEffectCoercive2008,paulPolarizationSwitchingEpitaxial2008,jiangEnablingUltralowvoltageSwitching2022}. 

Consequently, stable periodic poling in \BTO\ is highly sensitive to substrate-specific strain and boundary conditions and requires narrow, carefully engineered growth windows. Although electric field induced periodic domain inversion has recently been demonstrated on dysprosium scandate (DyScO\textsubscript{3}) substrates with strontium ruthenate (SrRuO\textsubscript{3}) electrodes \cite{aashnaPeriodicDomainInversion2024}, this remains a substrate-specific result of limited photonic relevance due to DyScO\textsubscript{3}’s high refractive index that does not allow mode confinement in the \BTO\ layer and losses in the SrRuO\textsubscript{3} electrode. Moreover, periodic poling can introduce additional domain walls and defects, increasing optical scattering losses in \BTO\ waveguides\cite{kimOriginNonabsorptiveScattering2025}.

These constraints motivate alternative $\boldsymbol{\chi}^{(2)}$ frequency conversion strategies, notably modal phase-matching, which avoids micro- or nanoscale periodic poling and ferroelectric domain engineering. In this strategy, the waveguide dimensions are chosen such that the effective refractive index of the pump mode at the fundamental frequency, typically the fundamental transverse electric (TE) or transverse magnetic (TM) mode, is equal to the effective refractive index of a second harmonic mode, typically a higher order mode. This results in waveguides which are generally simpler from a fabrication point of view, but which typically have lower efficiencies due to the mode order mismatch causing smaller mode overlaps. To combat this, the cross-section of the waveguides can be modified to help maximize the mode overlap. Comparable approaches have been shown in \LNO\ using double layered waveguides \cite{Du202505,Du202303,Hefti202510}, hybrid linear-nonlinear waveguides \cite{Honda202511,Luo201901,Wang202511}, or the introduction of asymmetries to the cross-section \cite{Zhang201306}. However, to date designs for maximizing the efficiency of frequency conversion processes with \BTO\ are missing. In this work, we propose hybrid \BTO-\TiO\ waveguides, in which a thin \TiO\ layer modifies modal confinement and significantly enhances nonlinear overlap \cite{Hsu2025,Evans2015}. Because this strategy relies solely on lithographic definition of the waveguide geometry, it is inherently scalable and avoids the limitations of domain inversion in \BTO.

The following sections develop a coupled-mode-theory design framework for these hybrid structures, quantify the achievable second-harmonic-generation efficiency, and benchmark their performance against both monolithic \BTO\ waveguides and established nonlinear photonic platforms. Our results show that these hybrid \BTO-\TiO\ waveguides provide a practical and CMOS-compatible pathway toward efficient  $\boldsymbol{\chi}^{(2)}$ devices on the emerging \BTO\ photonic platform.

\section{Working Principle of Hybrid \BTO\ Waveguides}

\begin{figure}[th]
\centering
\begingroup%
\includegraphics[]{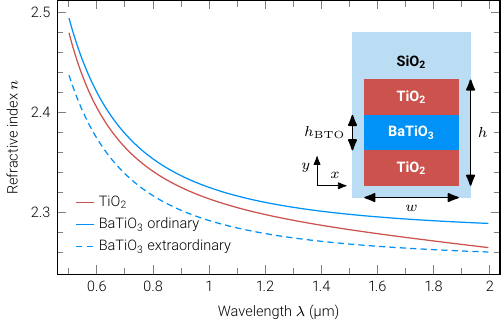}%
\endgroup%
\LREF\caption{Refractive indices of the two \BTO\ axes compared with \TiO. The inset shows the cross-section of the hybrid waveguide geometries discussed in this paper.}
\label{FIG:GEO}%
\end{figure}

The mode overlap between a fundamental frequency and second harmonic eigenmode in a waveguide can be calculated as \cite{Suhara2003},
\begin{equation}
	\kappa_m
    =
    \left\|
        \frac{2\omega\epsilon_0}{4}
        \iint
        \overline{\fun{\mathbf{E}_{m}}{x,y,2\omega}}
        \cdot
        \fun{\mathbf{P}_\mathrm{NL}}{x,y,2\omega}
        \dd x
        \,\dd y
    \right\|
\label{EQU:ModeOverlap}%
\end{equation}
\begin{equation}
	\fun{P_{\mathrm{NL},i}}{x,y,2\omega}
    =
    \sum_{jk}
    \frac{1}{2}
    \fun{\chi_{ijk}^{(2)}}{x,y}
    \fun{E_j}{x,y,\omega}
    \fun{E_k}{x,y,\omega}
\end{equation}
where $\kappa_m$ is the mode overlap, $\fun{\mathbf{E}_{m}}{2\omega}$ is the normalized electric field of the second harmonic eigenmode of interest, and $\fun{\mathbf{E}}{\omega}$ is the normalized electric field of the fundamental frequency input eigenmode. The units of $\kappa_m$ are $\frac{\sqrt{\text{W}}}{\text{m}}$, and therefore it represents how much energy is converted into the second harmonic mode per unit time and unit of propagation length, without considering potential phase-matching problems.

We note that in calculating the mode overlap, the fundamental frequency mode gives rise to the nonlinear polarization density field (mediated by $\boldsymbol{\chi}^{(2)}$), which in turn is projected on to the second harmonic mode and then integrated across the full waveguide cross-section. The second harmonic modes which are naturally phase-matched in monolithic cuboid waveguides tend to be higher order modes, meaning the electric fields have multiple lobes in the transverse plane, with alternating signs. This means that the integral across the full plane has both positive and negative contributions, leading to a smaller mode overlap. Hybrid waveguides can help solve this problem by selectively placing linear (centrosymmetric such that $\boldsymbol{\chi}^{(2)}=\mathbf{0}$) materials in the waveguide cross-section, so that the nonlinear interaction only takes place within lobes of a single sign.

\section{\TiO-\BTO-\TiO\ Platform Concept}
\zcref[]{FIG:GEO} shows how two established waveguide materials, \BTO\ and \TiO, can be combined for a hybrid waveguide design. \TiO\ would be a good candidate as the linear material to combine with \BTO\ for hybrid waveguides, as its refractive index has a value between the ordinary and extraordinary refractive indices of \BTO. This means that the modes supported by a hybrid waveguide core would be very similar to those for a monolithic \BTO\ core, since the mode profiles and effective refractive index only depend on the linear susceptibility. 

We propose to realize a \TiO–\BTO–\TiO\ multilayer stack using a layer-by-layer growth and bonding sequence that adds only minimal complexity to established oxide photonics fabrication workflows. In a first step, individual oxide layers are deposited and optimized separately: crystalline \BTO\ thin films are grown on a suitable growth substrate under conditions tailored for ferroelectric and nonlinear performance \cite{liEvolutionEpitaxialBaTiO32022,reynaudSiIntegratedBaTiO3ElectroOptic2023,mazetReviewMolecularBeam2015} or fabricated using crystal ion slicing and subsequent annealing\cite{izuharaSinglecrystalBariumTitanate2003,parkSingleCrystallineBaTiO32007,esfandiarStructuralOpticalProperties2025}, while \TiO\ layers are deposited independently using standard oxide thin-film techniques such as sputtering or atomic layer deposition, which are fully compatible with photonic device fabrication and known to yield low-loss waveguides with strong optical confinement\cite{bradleySubmicrometerwideAmorphousPolycrystalline2012,Fu202012,hegemanDevelopmentLowlossTiO22020a,Hayrinen2014}.

Subsequently, the \TiO\ and \BTO\ layers are combined via direct wafer bonding, exploiting the demonstrated compatibility of \BTO\ with metal-oxide layers such as \TiO\ and Al\textsubscript{2}O\textsubscript{3} \cite{Jin202009,abelLargePockelsEffect2019}. This bonding step enables the formation of a vertically symmetric \TiO–\BTO–\TiO\ stack, after which the original growth substrates can be removed if required. In a final step, standard lithography and dry-etching processes are applied to pattern the bonded stack into photonic structures such as waveguides or resonators. This decoupled growth–bonding–patterning sequence relaxes constraints imposed by direct epitaxial growth on photonic substrates, mitigates issues related to lattice mismatch and strain engineering, and provides a scalable and flexible route toward integrated \BTO-based electro-optic and nonlinear photonic devices.

\section{Geometry Optimization}
To design the hybrid waveguides, we first simulate monolithic \BTO\ waveguides to find geometries that naturally have good modal phase-matching with a second harmonic mode. The profile of this well phase-matched second harmonic mode can then be analyzed to see where the linear material (\TiO) should be placed to maximize the mode overlap. Then, simulations of this hybrid design can be performed to optimize the layer thicknesses and geometry, given the slight differences to the eigenmodes induced by the inclusion of the linear material.

\begin{figure*}[t]
\centering
\begingroup%
\phantomsubcaption\label{FIG:SHG:a}%
\phantomsubcaption\label{FIG:SHG:b}%
\phantomsubcaption\label{FIG:SHG:c}%
\phantomsubcaption\label{FIG:SHG:d}%
\includegraphics[]{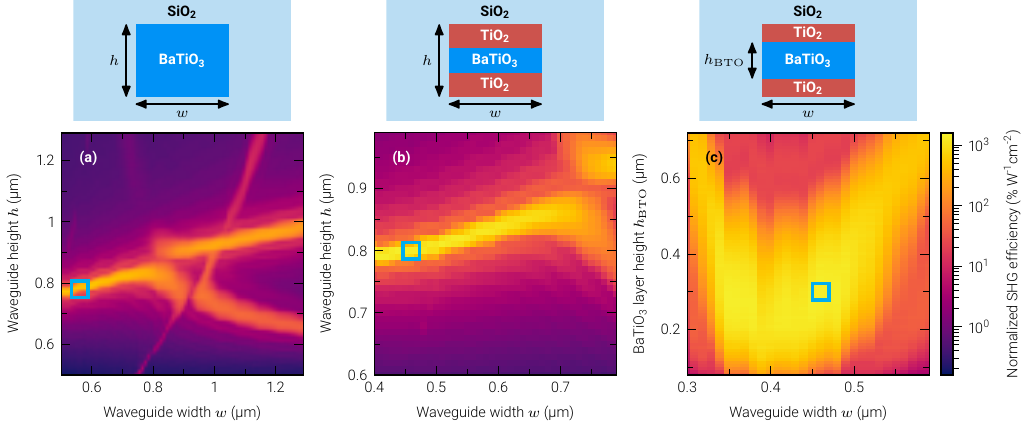}%
\endgroup%
\LREF\caption{Geometry optimization sweeps of the hybrid waveguides. \Lref{a} Monolithic \BTO\ waveguide cores showing bands of high efficiency where there are naturally second harmonic modes phase-matched to the $\text{TM}_{00}$ mode at the fundamental frequency. \Lref{b} Hybrid waveguides with varying total heights and widths, but all three layers kept equal. \Lref{c} Hybrid waveguides with the total core height fixed at 0.8 µm but varying widths and height of the \BTO\ layer. The cyan square in each plot represents the waveguide with the largest normalized SHG efficiency. Above each plot is the corresponding waveguide cross-section with the two varying geometric parameters indicated.}
\label{FIG:SHG}%
\end{figure*}

\zcref[]{FIG:SHG:a} shows simulation results of a monolithic \BTO\ cuboid waveguide of varying widths and heights, encased in \SiO. A coupled mode theory (CMT) based solving tool was used to simulate SHG in each waveguide, as this allows us to view exactly which eigenmodes are used in the interactions and where the energy is concentrated in the waveguide. For each waveguide geometry, the $\text{TM}_{00}$ mode at the fundamental wavelength (1550 nm) was given an initial amplitude of 1 $\sqrt{\text{W}}$, and all other mode amplitudes were set to 0. After a propagation distance of 100 µm, the normalized SHG efficiency of the waveguide was recorded.

Bands of efficient waveguides appear in the sweep, due to a particular second harmonic mode being well phase-matched to the input mode. The CMT simulation allows us to view the profile of these modes, and then selectively place a linear material to help improve the mode overlap, and thus SHG efficiency. For example, from \zcref{FIG:SHG:a} the most efficient waveguide was when the \BTO\ core had a height of 0.78 µm and a width of 0.56 µm (as indicated by the cyan square). In this case, the phase-matched second harmonic mode carrying the energy is the $\text{TE}_{02}$ mode, which has three main lobes stacked vertically in its electric field. A hybrid waveguide can then be composed by having sections of \TiO\ covering the top and bottom lobes, and \BTO\ in the central lobe. The \BTO\ layer is placed in the middle as this is where the input mode is concentrated.

\zcref{FIG:SHG:b} shows how the hybrid waveguide geometry was then optimized, first by keeping the height of all three layers equal but changing the total height and width of the waveguide. The bands of efficient structures follow a very similar shape as those from the monolithic \BTO\ waveguides, proving that the addition of the \TiO\ layers has little effect on the phase-matching. The ideal height for the total waveguide structure was found to be 0.8 µm. Then, in \zcref{FIG:SHG:c} the geometry was further optimized by keeping the total waveguide height fixed, but varying the height of the \BTO\ layer along with the width of the waveguide. The optimized hybrid waveguide was found to have a width of 0.46 µm, with a \BTO\ layer 0.30 µm high sandwiched between two layers of \TiO\, each 0.25 µm high.

The bands of high efficiency in \zcref{FIG:SHG:b} have a manufacturing tolerance of 30-40 nm, while in \zcref{FIG:SHG:c} the manufacturing tolerance of the \BTO\ layer height is $\sim$200 nm. This makes sense as these bands are determined by the phase-matching of a second harmonic mode with the input mode. Since \TiO\ has a refractive index very close to \BTO, the total structure height and width have a bigger impact on the effective refractive indices of the eigenmodes compared to the height of the individual layers in the core. This is advantageous from a manufacturing perspective, as it does not impose stringent thickness tolerances on the \BTO\ film itself. Instead, once the \BTO\ layer has been fabricated and its thickness measured, the \TiO\ cladding layers can be deposited accordingly to achieve a total stack height of 0.80 µm.

\begin{figure}[t]
\centering
\begingroup%
\phantomsubcaption\label{FIG:MODES:a}%
\phantomsubcaption\label{FIG:MODES:b}%
\phantomsubcaption\label{FIG:MODES:c}%
\phantomsubcaption\label{FIG:MODES:d}%
\phantomsubcaption\label{FIG:MODES:e}%
\phantomsubcaption\label{FIG:MODES:f}%
\includegraphics[]{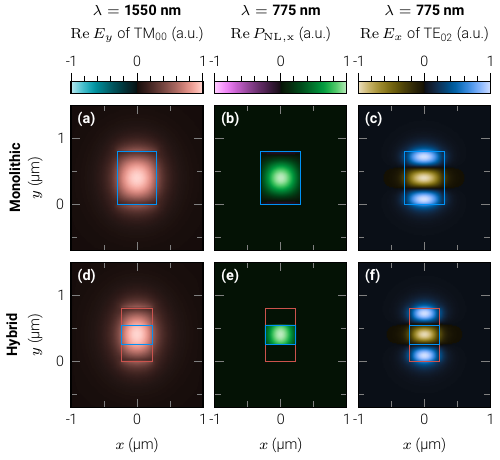}%
\endgroup%
\LREF\caption{Dominant components of the \Lref{a,d} fundamental electric, \Lref{b,e} nonlinear polarization density, and \Lref{c,f} second harmonic electric fields for \Lref{a,b,c} monolithic and \Lref{d,e,f} hybrid waveguide cores. The regions filled with \BTO\ and \TiO\ are outlined by blue and red lines, respectively. The remaining regions extending to the edges of each plot are filled with \SiO.}
\label{FIG:MODES}%
\end{figure}

\zcref{FIG:MODES} shows how the hybrid waveguides improve the mode overlap in the waveguide. The integral across the \BTO\ cross-section in the monolithic waveguides includes components from both signs in the second harmonic mode (\zcref{FIG:MODES:c}). In comparison, \zcref{FIG:MODES:f} shows how the hybrid structure isolates each lobe of the eigenmode to a specific layer in the waveguide core, and therefore the integral over the \BTO\ layer only includes a single sign. The result is that the mode overlap of the hybrid waveguide is 1.6 times larger, despite having a cross-sectional area with only 0.28 times as much \BTO\ compared to the monolithic waveguide.

\begin{figure}[h]
\centering
\begingroup%
\includegraphics[]{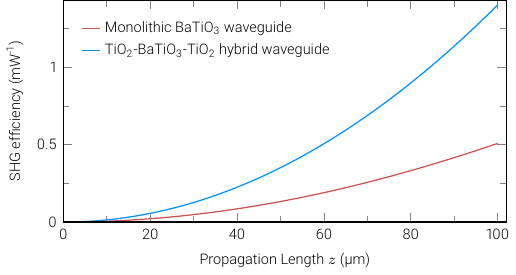}%
\endgroup%
\LREF\caption{Comparison of SHG efficiency between the monolithic \BTO\ waveguide and the hybrid waveguide over a 100 µm long propagation distance.}
\label{FIG:SHG_EFF_COMP}%
\end{figure}

This increased mode overlap also results in an increased SHG efficiency, as seen in \zcref{FIG:SHG_EFF_COMP} plotted over a propagation distance of 100 µm. The normalized SHG efficiency of the monolithic waveguide is 510.5 $\%\cdot\text{W}^{-1}\cdot\text{cm}^{-2}$, while for the hybrid structure it is 1404.5 $\%\cdot\text{W}^{-1}\cdot\text{cm}^{-2}$. This efficiency is significantly larger than theoretical efficiencies predicted for other material platforms such as monolithic GaP waveguides (6.1 $\%\cdot\text{W}^{-1}\cdot\text{cm}^{-2}$ \cite{anthurSHGinGAP}), or AlN-SiN hybrid waveguides (12 $\%\cdot\text{W}^{-1}\cdot\text{cm}^{-2}$ \cite{Honda202511}). As well, these designs show that \BTO\ waveguides can have theoretical efficiencies of the same order of magnitude seen in hybrid \LNO\ waveguides (2900 $\%\cdot\text{W}^{-1}\cdot\text{cm}^{-2}$ \cite{Luo201901}), periodically poled \LNO\ waveguides (3000 $\%\cdot\text{W}^{-1}\cdot\text{cm}^{-2}$ \cite{Wang201811}) and in particular periodically poled thin-film \LNO\ waveguides (1600 $\%\cdot\text{W}^{-1}\cdot\text{cm}^{-2}$ \cite{ChangThinFilm}), which is the standard implementation for \LNO-based integrated optics. It is also worth noting that because our hybrid waveguide platform uses modal phase-matching and therefore has an unchanging cross-section, these devices should be easier to fabricate compared to designs relying on periodic poling, leading to them being more reproducible, scalable, and capable of broadband operation.

\section{Conclusion}
In this work, we proposed a new platform and design scheme to increase the efficiency of nonlinear frequency conversion processes in \BTO\ waveguides. This was done by first using modal phase-matching to find geometries which naturally have high efficiencies without the need for periodic poling. By knowing the profile of the phase-matched modes, the nonlinear efficiency could be further enhanced by creating a linear-nonlinear hybrid structure, and selectively placing the nonlinear material in the waveguide cross-section.

We propose that this is best done by having a \TiO\ layer both above and below the \BTO\ layer, as these materials have similar linear refractive indices, and thus the phase-matching requirements are similar to a monolithic \BTO\ waveguide. Through optimization of the hybrid waveguide geometries, the normalized SHG efficiency was found to be 2.75 times larger than optimized monolithic \BTO\ waveguides, and of the same order of magnitude as various waveguide designs from \LNO. Overall, these hybrid waveguides offer the ability to introduce efficient and easily scalable nonlinear frequency conversion applications to the already established electro-optic modulation \BTO\ platform, opening the door to new possibilities for nonlinear and quantum photonic integrated circuits.

\section*{Methods}
\subsection*{Simulation Methods}
From Maxwell's curl equations the vector wave equation can be derived, the solutions of which are a series of eigenvectors with corresponding eigenvalues. Each eigenvector describes how an electric and magnetic field can propagate in the waveguide, and the corresponding eigenvalue is known as the propagation constant. The combined electric and magnetic field is known as an eigenmode, and once normalized, eigenmodes of a specific frequency have the unique property that they are all orthonormal to each other. Therefore, any electromagnetic field propagating through a waveguide can be decomposed into a linear superposition of eigenmodes as
\begin{align}
    \fun{\mathbf{E}}{x,y,z,\omega}
    &=
    \sum_{m}
    \fun{A_m}{z}
    \fun{\mathbf{E}_m}{x,y,\omega}
    \exp[0]{\ii \beta_m z}
\label{EQU:EigenmodeExpansionE}%
\\
    \fun{\mathbf{H}}{x,y,z,\omega}
    &=
    \sum_{m}
    \fun{A_m}{z}
    \fun{\mathbf{H}_m}{x,y,\omega}
    \exp[0]{\ii \beta_m z}
\label{EQU:EigenmodeExpansionH}%
\end{align}
where $A_m$ is the amplitude of the eigenmode, $\mathbf{E}_m$ and $\mathbf{H}_m$ are the electric and magnetic mode profiles in the transverse plane, $\beta_m$ is the propagation constant of the mode, and $z$ is the current position in the propagation direction.

Using \zcref{EQU:EigenmodeExpansionE} as an ansatz in the vector wave equation, we find a set of coupled differential equations that describe how each mode amplitude evolves as the mode propagates along the waveguide \cite{Suhara2003}:
\begin{multline}
    \frac{\partial}{\partial z}
    \fun{A_m}{z}
    =
    \ii \frac{\omega}{4} \epsilon_0
    \exp[0]{-\ii\beta_m z}
\\
    \int_{\mathbb{R}^2}
        \overline{\fun{\mathbf{E}_m}{x,y,\omega}}
        \cdot
        \fun{\mathbf{P}_\mathrm{NL}}{x,y,z,\omega}
    \,\dd x
    \,\dd y
\label{EQU:CMTEquation}%
\end{multline}
We then find the eigenmodes of a specific waveguide geometry and numerically solve the corresponding set of coupled differential equations from \zcref{EQU:CMTEquation} as an initial value problem. This results in a list of complex mode amplitudes, each evaluated at each $z$-step of the initial value solver. With these amplitudes, the total $\mathbf{E}$ and $\mathbf{H}$ fields at each frequency can be found in the waveguide using \zcref{EQU:EigenmodeExpansionE,EQU:EigenmodeExpansionH}.

If we assume that the electric field consists of only harmonic waves at the fundamental ($\omega$) and second harmonic ($2\omega$) frequencies, it can be expressed as,
\begin{align}
	\fun{\mathbf{E}}{t}
	&=
	\fun{\Re}{
		\fun{\mathbf{E}}{\mathbf{r},\omega}
		\exp[0]{-\ii \omega t}
		+
		\fun{\mathbf{E}}{\mathbf{r},2\omega}
		\exp[0]{-\ii 2\omega t}
	}
	\label{EQU:ElectricField}%
\end{align}
and the total guided power $\fun{P}{z}$ at a point $z$ in the waveguide can be decomposed into the contributions of the modes as
\begin{align}
    \fun{P}{z}
    &=
     \fun{P}{z,\omega}
    +\fun{P}{z,2\omega}
\\
    \fun{P}{z,q}
    &=
    \frac{1}{2}
    \int_{\mathbb{R}^2}
    \fun{\Re}{
        \fun{\mathbf{E}}{\mathbf{r},\omega}
        \times
        \overline{\fun{\mathbf{H}}{\mathbf{r},\omega}}
    }
    \cdot\hat{\mathbf{z}}
    \,\dd x
    \,\dd y
\\\nonumber
	&+
	\frac{1}{2}
	\int_{\mathbb{R}^2}
	\fun{\Re}{
		\fun{\mathbf{E}}{\mathbf{r},2\omega}
		\times
		\overline{\fun{\mathbf{H}}{\mathbf{r},2\omega}}
	}
	\cdot\hat{\mathbf{z}}
	\,\dd x
	\,\dd y
\end{align}
Without loss of generality, it can be assumed that the waveguide starts at $z=0$ and is excited there with a certain power at the fundamental frequency $\fun{P}{0,\omega}$. The normalized efficiency $\eta$ with which second harmonic photons are generated is then defined as
\begin{align}
    \fun{\eta}{z}
    &\coloneq
    \frac{
        \fun{P}{z,2\omega}
    }{
        \fun{P}{0,\omega}^2
    }
\label{EQU:NonlinearEfficiency}%
\end{align}
with units of $\left[\eta\right] = \text{W}^{-1}$. Since $\fun{P}{z,2\omega}$ is generally proportional to $\fun{P}{0,\omega}^2 z^2$, i.e.\ dependent on the square of the waveguide length, we can further define a normalized efficiency $\tilde{\eta}$
\begin{align}
    \tilde{\eta}
    &\coloneq
    \argmin_a
    \int_{0}^{\ell}
    \left[\fun{\eta}{z} - a z^2\right]^2
    \dd z
    =
    \frac{5}{\ell^5}
    \int_{0}^{\ell}
    z^2 \fun{\eta}{z}
    \dd z
\end{align}
with units of $\left[\eta\right] = \text{W}^{-1}\cdot\text{m}^{-2}$, where $\tilde{\eta}z^2$ fits the normalized efficiency $\fun{\eta}{z}$ in the least squares sense and $\ell$ is the actual length of the waveguide. Values of equivalent quantities are often given in literature in units of $\%\cdot\text{W}^{-1}\cdot\text{cm}^{-2}$.


\section*{Acknowledgments}
The authors acknowledge the Fraunhofer Attract Grant SILIQUA No.\ 40-04866, the BMBF projects MEXSIQUO Grant No.\ 13N16967 and SINNER Grant No.\ 16KIS1792, and the Collaborative Research Center (CRC/SFB) 1375 NOA.

\section*{Conflict of Interest}
The authors declare no conflict of interest.

\section*{Data Availability Statement}
The data that support the findings of this study are available from the corresponding author upon reasonable request.

\section*{Keywords}
Barium titanate, titanium dioxide, modal phase-matching, hybrid waveguide, nonlinear optics, integrated photonics

\printbibliography

\end{document}